\documentclass[%
superscriptaddress,
 amsmath,amssymb,
 aps,
 prx,
 twocolumn 
 ]{revtex4-2}

\usepackage{graphicx}
\usepackage{dcolumn}
\usepackage{bm}
\usepackage{xcolor}
\usepackage[caption=false]{subfig}
\usepackage{xfrac}
\usepackage{blindtext}
\usepackage{times}
\usepackage{comment}
\usepackage{hyperref}

\definecolor{darkblue}{HTML}{004D6B}
\definecolor{darkred}{HTML}{8c1515}
\hypersetup{
	colorlinks=true,
	urlcolor=darkblue,
	citecolor=darkred,
	linkcolor=darkblue,
	breaklinks
}

\begin{document}

\title{Unconventional Thermalization of a Three-Wave-Mixing Model}

\author{Evangelos Varvelis}
\affiliation{Institute for Complex Quantum Systems and IQST, Ulm University, 89069 Ulm, Germany}

\author{Miriam Resch}
\affiliation{Institute for Complex Quantum Systems and IQST, Ulm University, 89069 Ulm, Germany}

\author{Joachim Ankerhold}
\affiliation{Institute for Complex Quantum Systems and IQST, Ulm University, 89069 Ulm, Germany}

\date{\today}

\begin{abstract}

Understanding the boundaries between quantum thermalization and localization in many-body systems remains a central frontier of condensed matter and quantum information science. In this work, we investigate the dynamics and spectral properties of a generic model with long-range three-body-interaction, namely, a system with non-local three-wave-mixing. This model has been realized recently with a microwave Fabry-Perot cavity terminated on one end by a superconducting qubit mirror. Utilizing exact diagonalization techniques, we uncover a striking paradox: the global energy level spacing statistics show integrability, even though all dynamic observables and inverse participation ratios of the eigenstates indicate ergodicity and delocalization. We show that this behavior is a hallmark of strong Hilbert space fragmentation driven by kinematic constraints rather than an explicit global symmetry. Inside these sectors, dynamics scramble rapidly, as evidenced by the out-of-time-ordered correlator (OTOC), while global transport is heavily bottlenecked, resulting in a logarithmic relaxation to equilibrium. This picture is further confirmed by fluctuations in eigenstate entanglement entropy at the same energy. Finally, we demonstrate that the late time OTOC average scales with system size, providing a distinct experimentally accessible signature of the underlying three-body kinetic bottlenecks.

\end{abstract}

\maketitle

\section{Introduction}

Interest in thermalization of quantum systems has risen in the recent years along with the increased interest in the emerging field of quantum technologies. Approaches vary from studying the thermalization of such devices and ways to keep it under control to ensure their proper operation \cite{Guo2021, Berke2022, Iadecola2023, Blais2023} all the way to inverting this perspective and using these devices as probes to study fundamental questions about the very nature of thermalization of quantum systems \cite{Martinis2017, Michailidis2019, Chen2021, Gong2021}. However, thermalization of quantum systems remains an active field of research, exemplified from instances of paradoxically identifying the same system as both as avoiding thermalization via many-body localization (MBL) \cite{Manucharyan2022} and as thermalizing \cite{Pekola2024}.
Here we will focus on a generic model with three-body-interaction in order to unify these two opposing views. This in turn will shed light on the complexity of quantum many-body thermalization and the information encoded in various figures of merit to characterize it.  

The model under consideration is motivated by an experimental realization pioneered in \cite{Manucharyan2022}, essentially consisting of a microwave Fabry-Perot cavity in form of a long transmission line with weakly coupled probe antenna on one end and a superconducting fluxonium qubit on the other. This system is operated in the \textit{superstrong} coupling regime of multimode cavity quantum electrodynamics (cQED), i.e.\ the vacuum Rabi frequency of the qubit is larger than the free spectral range of the resonator. As a result, the qubit and cavity modes exchange excitations faster than the light takes to traverse the cavity, thus the profile of the cavity modes becomes dependent on the state of the qubit. Consequently, an effective description of the system replaces the bare cavity modes near resonance with the qubit with dressed modes arising from the qubit-cavity interaction. However, higher excitations of the further off-resonant lower frequency modes may still couple with the qubit via down-conversion processes, and by extension with the dressed photonic modes of the cavity. In fact, for superconducting qubit settings such as this, the conversion rate between near resonance dressed modes and many-particle excited off-resonant modes is in the order of a few MHz, well above the decoherence threshold. Beyond this specific realization, three-body interactions have been explored with ultracold atomic gases and they appear in effective theories for many-body nucleon systems, see e.g.\ \cite{hammer2013}.

A generic effective Hamiltonian in this mixed gauge of reservoirs of low $b_k$ and high frequency $a_k$ modes, respectively,  is the three-wave-mixing (TWM) Hamiltonian
\begin{align}
    H &= H_0 + H_\text{int},\label{FullHamiltonian}\\
    H_{0} &= \sum_{k = 1}^{N_b} \omega_k b_k^{\dagger}b_k + \sum_{k = 1}^{N_a} \Omega_k a_k^{\dagger}a_k,\label{FreeHamiltonian}\\
    H_{\text{int}} &= g\sum_{k = 1}^{N_b}\sum_{i,j=1}^{N_a} \sqrt{k}\ e^{-\vert i - j\vert/\xi}(a_i^\dagger a_j b_k + \text{h.c.}),\label{IntHamiltonian}
\end{align}
in units of $\hbar = 1$, which we will use throughout the paper. For the sake of clarity, from now on we will refer to the high frequency modes as \textit{polaritonic} and the low frequency modes as \textit{bare}, motivated by the described experimental situation.  For the respective mode frequencies we consider the simplest setting, where 
\begin{align}
    \omega_{k} &= \omega_0 + (k-1)\Delta + \delta\omega_k,\label{BareFrequencies}\\
    \Omega_{k} &= \omega_{N_b} + k\Delta + \delta\Omega_k\label{PolaritonFrequencies}
\end{align}
are the frequency ladders of the two types of modes with gradient $\Delta$ and respective disorder shifts $\delta\omega_k$ and $\delta\Omega_k$. The latter reflect the fact that in actual realizations, in particular in the one mentioned above, frequencies of individual modes may vary from the ideal equidistant ladder either due to imperfections during fabrication or due to sluggish diffusion of system parameters. Note that $\omega_0$ simply sets the energy scale. We chose to set $\omega_0 = 5\Delta$ for all our calculations without loss of generality and  in approximate agreement with the reported values in \cite{Manucharyan2022}. Additionally, because $\omega_{N_b}$ is the highest bare mode frequency, in the absence of disorder, the lowest polaritonic mode would be equidistant to the last bare mode and the next higher polaritonic mode, making the distinction between the two somewhat arbitrary. The parameter $g$ is the coupling strength for the three-body process describing the interaction term. In the cited experiment, the latter depends directly on the qubit spontaneous emission linewidth and  we refer to \cite{Manucharyan2022} for further details.
The exponential coefficients in the mixing term ensure that the polaritonic modes can down-convert only to another polaritonic mode within a few (dimensionless) localization lengths $\xi$. By tuning $\xi$ one can sweep from a regime of localized interaction to a regime of long-range coupling in the energy sector. 

In the usual folklore of quantum thermalization, a closed quantum system thermalizes when it can act as a bath for its subsystems yielding expectation values of local observables in agreement with the microcanonical ensemble. One way for a quantum system to thermalize is described by the eigenstate thermalization hypothesis (ETH) \cite{Deutsch, Srednicki}. In practice that means that the time evolution of a section of the system equilibrates after some time and that the equilibrium state is a thermal state. However, this is also state dependent, and in fact there are combinations of initial states and systems that avoid thermalization completely by attaining memory of their initial state in the form of revivals, and with observable statistics in disagreement with the microcanonical ensemble. The usual manifestation of this behavior for interacting systems is either in the form of MBL \cite{Serbyn2013, Huse2014} for extended regions of the Hilbert space that avoid thermalization or in the form of quantum many-body scars (MBS) for small fragments of the Hilbert space around classical periodic trajectories that avoid thermalization. In recent years another alternative to thermalization has emerged in the form of Hilbert space fragmentation (HSF). These systems are characterized by emergent symmetries that fragment the Hilbert space in sectors that are effectively decoupled and thus exhibiting non-ergodic behavior. 

We used a comprehensive suite of diagnostics for distinguishing between thermal and non-thermal behavior for our system. Our first set of diagnostics was the usual probes employed in MBL to thermal transition studies, level-spacing statistics \cite{Serbyn2016}, inverse participation ratios (IPR) \cite{Mirlin2008} and eigenstate entanglement \cite{Moudgalya2022}. The first one is quantified via the Kullback-Leibler (KL) divergence of the adjacent level spacing ratio distributions from the Poisson and Gaussian orthogonal ensemble (GOE). Finally we also employed two dynamic observables: the odd-even polaritonic mode imbalance \cite{Altman2015} and out-of-time-order correlators (OTOC) \cite{Swingle2018, Benenti2021} to measure the scrambling potential of the system dynamics. We find that while the system experiences a localized to thermal transition, the thermal region is exhibiting unconventional behavior, in the sense that it violates ETH. We surmise that this is due to hidden kinetic constraints related to the localization length $\xi$ of the interaction term.

\section{Model Characteristics}
\label{sec:model-characteristics}

In our calculations throughout the paper we treat the disorder potentials $\delta\omega_k$ and $\delta\Omega_k$ as random shifts sampled from a uniform distribution in the interval $\left[-\frac{w \Delta}{2},\frac{w \Delta}{2}\right]$ where $w$ is the dimensionless parameter that we will refer to as \textit{disorder strength}. We will limit $0\leq w\leq 1$ in all our calculations. The physical motivation behind this is that the cavity modes are labeled such that $\omega_k \leq \omega_{k+1}$ and $\Omega_k \leq \Omega_{k+1}$. Allowing for $w > 1$ could reverse the order of modes which is however not physical, in the sense that we can map the system back into a $w\leq 1$ definition simply by relabeling the modes.

Furthermore, we note that the Hamiltonian of Eq.~\eqref{FullHamiltonian} has an exact symmetry, which is the conservation of the total polaritonic number of excitations $\hat{n}_a = \sum_{k=1}^{N_a} \hat{n}_{a,k} = \sum_{k=1}^{N_a} a_k^\dagger a_k$. This operator is an integral of motion since it commutes with the Hamiltonian and therefore allows us to work in fixed number of polariton excitations sectors. In such a sector the possible Fock space configurations of the polariton excitations are simply a stars-and-bars calculation of $n_a$ stars and $N_a - 1$ bars, yielding the total number of configurations $d_a = \frac{(n_a+N_a-1)!}{n_a!(N_a-1)!}$.

On the other hand, there is no apparent symmetry for the bare modes and their total excitation number is not conserved, which means that we need to truncate their energy levels. Since energy is preserved in the system a reasonable way to truncate is to consider the extreme scenario where all polariton excitations reside in the highest mode and cascade down to the lowest. In the disorder free case and assuming for the moment $\omega_0 = \Delta$ the energy released from this de-excitation would equal the energy to excite bare mode 1 $n_a(N_a-1)$ times, bare mode 2 $\lfloor n_a(N_a-1)/2\rfloor$ times and so on. For larger $\omega_0$ such as the one we will use here this is a conservative upper bound of how many levels are needed. Thus the total number of bare mode configurations we will work with is $d_b = \prod_{k=1}^{N_b}\left(\left\lfloor\frac{n_a(N_a-1)}{k}\right\rfloor + 1\right)$, and therefore the total dimension of the Hilbert space will be given by $d_\mathcal{H} = d_a d_b$. 

In this work, we also inherit the convention from \cite{Manucharyan2022} and fix $N_b = \lfloor N_m/3\rfloor$ with $N_m$ the total number of modes, bare and polaritonic, that we will consider. Thus the Hilbert space can be completely characterized by the total number of modes and polaritonic excitations. We also note that this Hamiltonian is completely integrable in the $\xi\rightarrow 0$ limit and it can be exactly diagonalized by the displacement transformation
\begin{equation}\label{DisplacementOperator}
    D_b = \exp\left[-g\hat{n}_a\sum_{k=1}^{N_b}\frac{\sqrt{k}}{\omega_k}(\hat{b}_k^\dagger - \hat{b}_k) \right].
\end{equation}
This is not true for any finite $\xi$ however.

\section{Spectral Statistics}

\begin{figure*}
    \centering
    \includegraphics[width=0.99\textwidth]{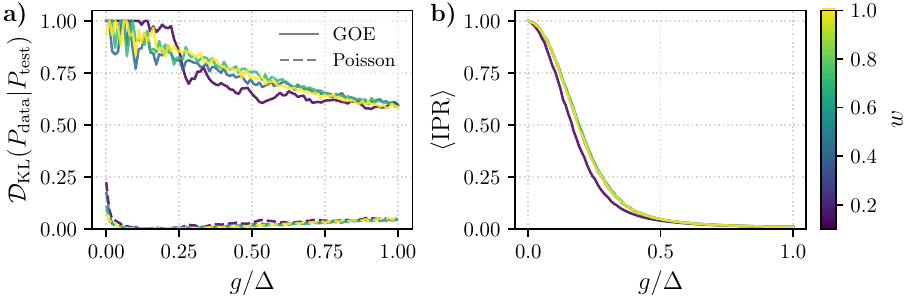}
    \caption{Spectral level spacing statistics and eigenstate localization. (a) Clipped Kullback-Leibler divergence $\min(\mathcal{D}_{\text{KL}}, 1)$ of the level-spacing $r$-ratio distribution [Eq.~\eqref{rRatios}] relative to GOE (solid) and Poisson statistics (dashed) [Eqs.~\eqref{rRatiosGOEDis} and \eqref{rRatiosPoissonDis}] as a function of coupling strength $g/\Delta$. Values are clipped at 1 for visual clarity as larger divergences indicate saturated statistical distance. (b) Disorder and spectral averaged Inverse Participation Ratio $\langle \text{IPR} \rangle$ [Eq.~\eqref{IPRDefinition}] versus $g/\Delta$. Curves for disorder strengths $w > 0.1$ collapse onto a nearly identical profile. Colorbar indicates disorder strengths $w \in \{0.1, 0.4, 0.7, 1.0\}$. All spectra are obtained via exact diagonalization of Eq.~\eqref{FullHamiltonian} for a system of 7 total modes (5 polaritonic, 2 bare) with $n_a = 3$ excitations ($d_{\mathcal{H}} = 3185$), interaction localization length $\xi = 1$, and averaged across 100 realizations. To filter out edge effects, analysis is restricted to the central 80\% of eigenstates in each spectrum.}
    \label{fig:SpectralData}
\end{figure*}

A stark contrast between the spectra of ergodic and non-ergodic systems is the presence of level repulsions. Systems that are quantum chaotic and ergodic have spectra drawn from some random matrix Gaussian ensemble and exhibit avoided crossings between their energy levels with varying interaction strength. On the opposite side, systems that are non-ergodic and do not explore the Hilbert space significantly have uncorrelated spectra with energy levels that cross and spacing statistics that follow a Poisson distribution. 

We begin the investigation of our system's behavior by looking at the spectral statistics, specifically through the \textit{normalized adjacent level spacing ratios} 
\begin{equation}
    r_n = \frac{\min(E_{n+2} -  E_{n+1},E_{n+1} -  E_{n})}{\max(E_{n+2} -  E_{n+1},E_{n+1} -  E_{n})},\label{rRatios}
\end{equation}
where $E_n$ are the eigenenergies of the system labeled in an ascending order. The reason behind choosing the definition in Eq.~\eqref{rRatios} and not the level spacings directly is in order to avoid the tedious procedure of spectral unfolding \cite{rRatioHuse}. For the $r$ ratios the GOE distribution \cite{Giraud2013}, which is the relevant Gaussian ensemble for our system, is 
\begin{equation}
    P_\text{GOE}(r) = \frac{27}{4}\frac{r+r^2}{(1+r+r^2)^{5/2}},\label{rRatiosGOEDis}
\end{equation}
while the Poisson distribution takes the form
\begin{equation}
    P_\text{P}(r) = \frac{2}{(1+r)^2}.\label{rRatiosPoissonDis}
\end{equation}
 
Since we wish to monitor the transition between the two limiting statistics, the systematic way to quantify the similarity between the calculated $r$ distribution of Hamiltonian Eq.~\eqref{FullHamiltonian} with the theoretical expressions Eqs.~\eqref{rRatiosGOEDis} and \eqref{rRatiosPoissonDis}  for each coupling and disorder strength is the Kullback-Leibler divergence 
\begin{equation}
    \mathcal{D}_\text{KL}(P\vert Q) = \sum_n P(n)\log\left(\frac{P(n)}{Q(n)}\right).\label{KullbackLeibler}
\end{equation}
The sum over $n$ refers to the binning of the two distributions which are compared. When the two distributions are identical the KL divergence vanishes while the greater their mismatch the higher the value of KL. We normalize the KL divergences such that the distance between the pure Poisson and GOE distributions is unity in each respective direction $\mathcal{D}_\text{KL}(P_\text{P}\vert P_\text{GOE}) = \mathcal{D}_\text{KL}(P_\text{GOE}\vert P_\text{P}) = 1$. Notice that KL divergence is not symmetric in its arguments and thus the divergence of the calculated distribution of the Hamiltonian in Eq.~\eqref{FullHamiltonian} from either of the two theoretical distributions has to be scaled differently. 

Usually the distinction between ergodic and non-ergodic is also reflected by the wavefunctions of the eigenstates, with ergodic systems having eigenstates whose wavefunctions are extended throughout the Hilbert space while for non-ergodic systems the states are more localized to a smaller subspace. This distinction can be quantified by the average inverse participation ratio 
\begin{equation}
    \langle\text{IPR}(g)\rangle_{\Delta E} = \frac{1}{N_{\Delta E}}\sum_{m=1}^{N_{\Delta E}}\sum_{n=1}^{d_\mathcal{H}} \vert\langle E_{n}(0) \vert E_{m}(g)\rangle\vert^4,\label{IPRDefinition}
\end{equation}
where $\vert E_n(g)\rangle$ are the eigenstates of the interacting Hamiltonian Eq.~\eqref{FullHamiltonian} while $\vert E_n(0)\rangle$ are the eigenstates of the free Hamiltonian Eq.~\eqref{FreeHamiltonian}. $N_{\Delta E}$ is the number of eigenstates that we average over in an energy window around the center of the bulk of the spectrum. For any system average IPR can vary from 1 (fully localized) all the way down to $1/d_\mathcal{H}$ (fully delocalized).

At interaction localization length $\xi = 1$ and for all system sizes available to us numerically, KL divergence exhibits the same behavior, for any finite disorder strength $w$, the $r$ ratio distribution seems to follow the Poisson distribution much more closely (see Fig.~\ref{fig:SpectralData}(a)) even as the IPR has already started to indicate a transition to delocalization (see Fig.~\ref{fig:SpectralData}(b)). Even though for increasing $g/\Delta$ ratio the distribution becomes more GOE-like, as evidenced by the slight drop of KL divergence from GOE in Fig.~\ref{fig:SpectralData}(a), it never comes close enough to be identified as GOE.

The IPR has the main disadvantage that it is a basis dependent measure. Specifically as defined in Eq.~\eqref{IPRDefinition} this is a measure of the spread of the wavefunctions of the eigenstates in the eigenbasis of the free Hamiltonian Eq.~\eqref{FreeHamiltonian}, however there is nothing special about that choice. Additionally, in recent years numerous studies have explored the case of non-ergodic delocalization \cite{Tang2022, Altshuler2020, Faoro2019, Monthus2017}, a phenomenon that highlights a critical limitation of the strength of the IPR as a measure of ergodicity. To reinforce and expand our findings avoiding the issues of the IPR we turned towards dynamical probes. 

\section{Dynamical Observables}

\begin{figure*}
    \centering
    \includegraphics[width=0.99\textwidth]{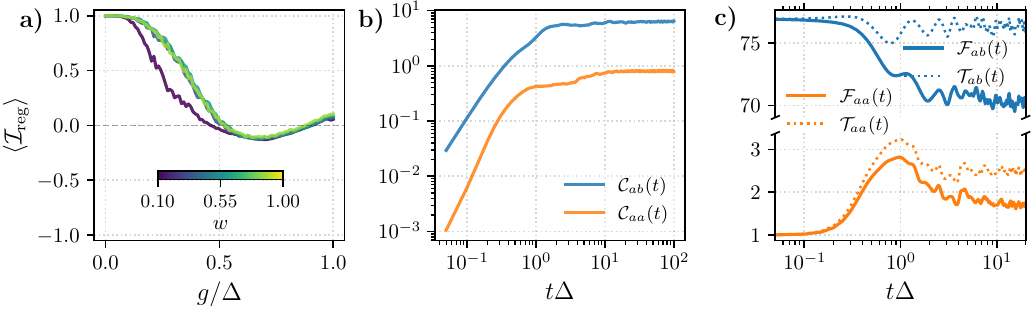}
    \caption{Imbalance relaxation and real-time operator scrambling. (a) Disorder-averaged regularized imbalance $\langle \mathcal{I}_{\text{reg}} \rangle$ [Eq.~\eqref{RegularisedImbalance}] versus coupling strength $g/\Delta$ for varying disorder strengths $w \in \lbrace 0.1, 0.4, 0.7, 1.0\rbrace$ (inset colorbar), averaged over 100 realizations. System size: 7 total modes (5 polaritonic, 2 bare) with $n_a = 3$ excitations ($d_\mathcal{H} = 3185$). Curves for $w > 0.1$ collapse onto a nearly universal profile. (b) Time dynamics of OTOCs $\mathcal{C}_{ab}(t)$ and $\mathcal{C}_{aa}(t)$ [Eq.~\eqref{OTOCDef}] for a single disorder realization with $w = 0.5$ and coupling $g/\Delta = 2.0$. System size: 6 total modes (4 polaritonic, 2 bare) with $n_a = 3$ excitations ($d_\mathcal{H} = 1000$). (c) Decomposition of the OTOCs into four-point $\mathcal{F}(t)$ (solid curves) and two-point $\mathcal{T}(t)$ correlator contributions (dotted curves) [Eqs.~\eqref{OTOCDef}--\eqref{FourPointOTOCDef}] for the same single realization as in (b). In (b) and (c), blue curves ($ab$) denote correlators between the highest-frequency polaritonic mode $\hat{n}_{a,N_a}$ and lowest-frequency bare mode $\hat{n}_{b,1}$, while orange curves ($aa$) correspond to correlators between highest- and lowest-frequency polaritonic modes $\hat{n}_{a,N_a}$ and $\hat{n}_{a,1}$ respectively. All time evolutions are obtained via exact diagonalization of Eq.~\eqref{FullHamiltonian} for localization length $\xi = 1$.}
    \label{fig:Imbalance_and_OTOC}
\end{figure*}

Since by definition thermalization of a system is a statement about the long time dynamics of the system, a study of thermalization for our system would be incomplete without the inclusion of dynamical observables. We used two diagnostics: the imbalance and the OTOC.

The imbalance, used in many studies of quantum thermalization \cite{Kohlert2023,Scherg2021}, is a direct test of the compliance of the long time behavior of the system with the predictions of the appropriate thermodynamic ensemble. Specifically for our system we define it as 
\begin{equation}\label{Imbalance}
    \mathcal{I}(t) = \frac{\langle\hat{n}_{a,\text{odd}}(t)\rangle - \langle\hat{n}_{a,\text{even}}(t)\rangle}{n_a},
\end{equation}
where $n_a$ is the constant total number of polaritonic excitations in the system, the first term in the numerator is the time evolved expectation value of the total polaritonic excitations on an odd index mode while the second term is the respective quantity for even index modes. When all excitations are gathered on odd or even modes the imbalance obtains an extremal value of $+1$ and $-1$ respectively. Conversely if the excitations are uniformly distributed the imbalance vanishes. We account only for the occupation numbers of polaritonic modes because they are conserved in total and only then can the imbalance have a meaningful interpretation. Since a system that does not thermalize attains memory of its initial state for very long times, in order to differentiate between this and ergodic behavior we need to initialize the system in a state far from equilibrium that extremizes imbalance.

This generic argument about how the value of the imbalance works as a thermalization witness has been repeated many times in the literature. However, there are some underlying assumptions about this that actually do not hold for our study. The generic statement that the imbalance Eq.~\eqref{Imbalance} vanishes for a thermalizing system is true assuming that one takes the thermodynamic limit and the $t\rightarrow\infty$ limit, and furthermore that the system has translational invariance. For the finite size system with an energy gradient that we study here the violation of these assumption leads to significant deviations for the actual thermal expectation value of the imbalance as defined in Eq.~\eqref{Imbalance}. Therefore we need to define what the thermal value of imbalance is in our case. 

First of all, we define the effective inverse temperature $\beta_\text{eff}$ of the initial state $\vert \psi_0\rangle$ from the equation $\langle\psi_0\vert H\vert\psi_0\rangle = \frac{1}{\mathcal{Z}}\text{Tr}\left(H e^{-\beta_\text{eff}H}\right)$ where $\mathcal{Z} = \text{Tr}\left(e^{-\beta_\text{eff}H}\right)$ is the partition function. Due to quantum typicality most choices of initial state land somewhere in the middle of the energy landscape and therefore correspond to some high temperature. This is also true for our system however we will not approximate this as $\beta_\text{eff} = 0$ and include the finite correction that the actual temperature induces. The thermal expectation value of the imbalance in that temperature is simply
\begin{equation}\label{ThermalImbalance}
    \langle\hat{\mathcal{I}}\rangle_\text{th} = \frac{\text{Tr}\left(\hat{\mathcal{I}} e^{-\beta_\text{eff}H}\right)}{\mathcal{Z}},
\end{equation}
where we used the operator form of the imbalance $\hat{\mathcal{I}} = (\hat{n}_{a,\text{odd}} - \hat{n}_{a,\text{even}})/n_a$. If the system is ergodic then the long time average of the imbalance
\begin{equation}\label{LongTimeAverageImbalance}
    \overline{\mathcal{I}} = \lim_{T\rightarrow\infty}\frac{1}{T}\int_{0}^{T}dt \mathcal{I}(t),
\end{equation}
should converge to the canonical ensemble prediction of Eq.~\eqref{ThermalImbalance}. In order to use the same reference values from the literature we define the \textit{finite size and temperature regularized imbalance} 
\begin{equation}\label{RegularisedImbalance}
    \mathcal{I}_\text{reg} = \frac{\overline{\mathcal{I}} - \langle\hat{\mathcal{I}}\rangle_\text{th}}{1-\text{sgn}\left(\overline{\mathcal{I}} - \langle\hat{\mathcal{I}}\rangle_\text{th}\right)\langle\hat{\mathcal{I}}\rangle_\text{th}},
\end{equation}
where sgn is the sign function. Notice that with this definition, when all excitations are on odd(even) modes $\overline{\mathcal{I}} = 1(-1)$ and therefore $\mathcal{I}_\text{reg} = 1(-1)$ as well, since $-1\leq\langle\hat{\mathcal{I}}\rangle_\text{th}\leq 1$, while for $\mathcal{I}_\text{reg} = 0$ we have true coincidence with the thermal expectation value.

We present the results for the regularised imbalance in Fig.~\ref{fig:Imbalance_and_OTOC}(a). Since we used the same system size and parameters the results are directly comparable with the ones in Fig.~\ref{fig:SpectralData}. We notice that the imbalance also tracks a phase change with increasing coupling strength similar to the IPR, but the spectrum of the system seems to be consistently Poisson across the phase change to thermal behavior the other two metrics indicate.

In order to tackle this apparent paradox we used the infinite temperature OTOC: $\mathcal{C}_{ij}(t) = -\text{Tr}\left(\left[\hat{n}_i(t),\hat{n}_j(0)\right]^2\right) / d_\mathcal{H}$, where $\hat{n}_i$ is the occupation number of mode $i$ and $d_\mathcal{H}$ the dimension of the Hilbert space. Since the operators are Hermitian and the time evolution unitary we can expand this expression to the following decomposition
\begin{align}
    \mathcal{C}_{ij}(t) &= \mathcal{T}_{ij}(t) - \mathcal{F}_{ij}(t),\label{OTOCDef}\\
    \mathcal{T}_{ij}(t) &= \frac{2}{d_\mathcal{H}}\text{Tr}\left(\hat{n}_i(t)^2\hat{n}_j(0)^2\right),\label{TwoPointOTOCDef}\\
    \mathcal{F}_{ij}(t) &= \frac{2}{d_\mathcal{H}}\text{Tr}\left(\hat{n}_i(t)\hat{n}_j(0)\hat{n}_i(t)\hat{n}_j(0)\right).\label{FourPointOTOCDef}
\end{align}
$\mathcal{C}_{ij}(t)$ is the full OTOC, and we will refer to the terms $\mathcal{T}_{ij}(t)$ and $\mathcal{F}_{ij}(t)$ as two and four point correlators respectively, the number primarily referring to how many times the term involves a time evolution. The reason for using the OTOC is that it can probe the dynamical signatures of thermalization independent from the choice of initial state. This unfortunately comes at the price that it is a much harder to use both experimentally \cite{Green2022} and theoretically. The main bottleneck for both is that these types of correlators involve time evolving the system, perturbing locally and then rewinding the time evolution, particularly the four-point correlator in Eq.~\eqref{FourPointOTOCDef}. The rewinding is experimentally tricky to achieve and computationally expensive to simulate.

This complication requires to restrict our choice of OTOCs to investigate in order to get a complete picture of our system without calculating the full spectrum of $\mathcal{C}_{ij}(t)$ coefficients. We chose to focus on two OTOCs, one that is between the occupation number of the highest frequency polaritonic mode $\hat{n}_{a,N_a}$ and the lowest frequency bare mode $\hat{n}_{b,1}$ which we refer to as $\mathcal{C}_{ab}(t)$ and one that is between the occupation number of the highest frequency polaritonic mode $\hat{n}_{a,N_a}$ and the lowest frequency polaritonic mode $\hat{n}_{a,1}$ which we refer to as $\mathcal{C}_{aa}(t)$. These already give a detailed picture about system dynamics. On one hand the $ab$ OTOC is describing how the polaritonic modes interact with the bare modes and on the other the $aa$ OTOC is describing how the polaritonic modes interact with themselves. Furthermore, the $ab$ related modes are nearest neighbor in the sense that the exponential amplitude with distance in the interaction term of Eq.~\eqref{IntHamiltonian} makes the polaritonic modes $N_a$ and $N_a - 1$ nearest neighbor and they couple mostly via the lowest bare mode since this is the one that violates energy conservation the least. In contrast the $aa$ OTOC is between the furthest away possible modes including information about the dependence of the OTOC with distance.

Regardless of the added complexity, the OTOC probes the thermalization properties and the information scrambling of the dynamics of a system in greater detail. The time evolution of the OTOC can be distinguished in three domains. The first one is the \textit{early time} \cite{Garcia2018} domain which describes the scrambling speed of information in the system. At $t = 0$, the two observables are space-like separated, meaning that they commute, and therefore the OTOC is zero. However, as time evolves the initial perturbation starts to evolve and explore the available degrees of freedom rapidly. The OTOC therefore rises, defining a quantum information light cone. The rise can be exponential for many systems with chaotic classical counterpart \cite{Rozenbaum2017,Chavez2019} or power law for systems without one \cite{Craps2020} and logarithmic for MBL. For finite systems, this domain ends at a time called the \textit{scrambling time}, after the system has explored the entire Hilbert space available. 

For the example case presented in Fig.~\ref{fig:Imbalance_and_OTOC}(b), since this is plotted in log-log scale it is clear that the scrambling of our system is described by a power-law and the scrambling time is of the order of $\approx 0.4/\Delta$. Even though the behavior of the system during scrambling time is known not to be universal, it is interesting to note that this power-law behavior, typically associated with systems lacking a direct semi-classical limit, characterizes our model despite its connection to classical parametric down-conversion. We also notice that the scrambling time is very brief, especially when compared to the other time scales of the OTOC which we will mention shortly. We surmise that this is most likely due to the fact that while the exponential weights in the coupling do quasi-localize the interaction, in reality the system maintains an all-to-all type of connectivity allowing for multiple channels of communication between two modes.

The next time regime is called the \textit{intermediate times} and it is by far the least understood domain of the three. It is generally shorter and directly after the scrambling time. At this stage, after the system has explored itself fully, it relaxes to equilibrium, witnessed primarily through the decay of the the four-point correlator of Eq.~\eqref{FourPointOTOCDef}, and the approach of the full OTOC to a plateau. Note that the full OTOC is still rising in this interval. For chaotic systems this decay is of the form of Pollicott-Ruelle resonances \cite{Venegeroles2008, Garcia2018} and it is an exponential decay. This is not however the case for our system, as evident from the log-linear plot in Fig.~\ref{fig:Imbalance_and_OTOC}(c). Our system seems to approach equilibrium with a logarithmic decay. Our second piece of information from the OTOC is therefore that the system exhibits a glass-like or constrained relaxation mechanism that slows-down equilibration. We also note that the equilibration time, meaning the duration between the scrambling time and the start of a plateau, is $\approx 3.6/\Delta$, an order of magnitude larger than the scrambling time.

\begin{figure*}
    \centering
    \includegraphics[width=0.99\textwidth]{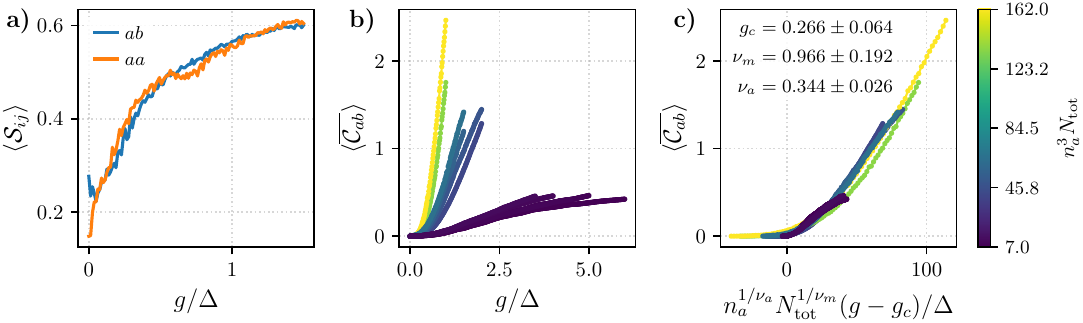}
    \caption{Scrambling dynamics and finite-size scaling. (a) Disorder-averaged spectral entropy $\langle S_{ij} \rangle$ [Eq.~\eqref{SpectralEntropy}] as a function of the coupling strength $g/\Delta$ for a system of 8 total modes (6 polaritonic, 2 bare) with $n_a = 2$ polaritonic excitations ($d_{\mathcal{H}} = 1386$), averaged over 100 realizations. Curves depict cross-sector correlators between the highest-frequency polaritonic mode and either the lowest-frequency bare mode ($ab$, blue) or lowest-frequency polaritonic mode ($aa$, orange). (b) Long-time average of the $ab$ OTOC $\langle \overline{\mathcal{C}_{ab}} \rangle$ [Eq.~\eqref{OTOCLongTimeAverage}] versus $g/\Delta$ across varying system sizes ($n_a \le 3$, $70 \le d_{\mathcal{H}} \le 1386$). (c) Data collapse of the OTOC curves using the finite-size scaling variable $n_a^{1/\nu_a} N_{\text{tot}}^{1/\nu_m}(g - g_c)/\Delta$, with critical coupling $g_c = (0.266 \pm 0.064)\Delta$ and critical exponents $\nu_m = 0.966 \pm 0.192$ and $\nu_a = 0.344 \pm 0.026$. Colorbar indicates the system scale parameter $n_a^3 N_{\text{tot}}$. All time evolutions are computed via exact diagonalization of Eq.~\eqref{FullHamiltonian} for interaction localization length $\xi = 1$ and disorder strength $w = 0.5$, with realization count $N_r$ dynamically chosen such that $N_r \cdot d_{\mathcal{H}} \approx 7 \times 10^4$ ($N_r \ge 100$).}
    \label{fig:OTOC_Scaling_and_Entropy}
\end{figure*}

Finally, the \textit{late time} \cite{Garcia2018} regime is characterized by the equilibration of the OTOC, witnessed by a plateau in the time average with potentially small oscillations around the mean. This part has also been extensively studied in the literature, and it supports two robust indicators of thermalization. The first one is concerned with the nature of the oscillations around the equilibrated mean. For an MBL system or more generally a system that attains memory of its initial state and avoids thermalization, these late time oscillations have the appearance of beating signals while a thermalizing system will exhibit oscillations more reminiscent of white noise. 

To quantify this difference in the quality of the oscillations we used the spectral entropy of the late time OTOC
\begin{equation}\label{SpectralEntropy}
	\mathcal{S}_{ij} = -\frac{1}{\log_2(N_t/2)}\sum_{n=1}^{N_t/2} \vert\tilde{\mathcal{C}}_{ij}(\omega_n)\vert^2\log_2 \vert\tilde{\mathcal{C}}_{ij}(\omega_n)\vert^2
\end{equation}
where $\tilde{\mathcal{C}}_{ij}(\omega_n)$ is the normalized discrete Fourier transformed late time OTOC evaluated at frequency $\omega_n$ and $N_t$ is the number of time samples of the original signal for the OTOC. The reason for using half of $N_t$ in this expression is because our Hamiltonian is real and symmetric and therefore so is the OTOC, thus the Fourier transform is symmetric around the DC component, which has been subtracted from the time signal of the OTOC. To understand how the spectral entropy quantifies the quality of oscillations, consider the two extreme cases: If the OTOC signal oscillates with a single dominant frequency and because the spectral components $\tilde{\mathcal{C}}_{ij}(\omega_n)$ are normalized the spectral entropy vanishes. On the contrary when the probed signal is white noise the spectral entropy takes the maximum value of $1$. This is a similar diagnostic to the inverse participation ratio of the Fourier transformed OTOC proposed in \cite{Fortes2019}. In fact from our calculation we observed that it is roughly a mirror image of the spectral entropy around the value $1/2$ since their extremal values of $0$ and $1$ correspond to the opposite frequency distributions.

We present the disorder averaged spectral entropy for a system of 8 modes in total, 6 polaritonic and 2 bare, with $n_a = 2$ excitations in total in the polaritonic sector in Fig.~\ref{fig:OTOC_Scaling_and_Entropy}(a). Since from all of the previous plots it has become apparent that the disorder strength $w$ does not play a significant role as long as it is not $w\lesssim 0.1$ from now on we fix its value at $w = 0.5$. It is clear that, while the value of the entropy does not reach either of the two extreme values, the oscillations of the OTOC exhibit a departure from beats of a few frequencies to a much denser spectrum as the interaction strength increases. The fact that the entropy increase seems to slow down near the end of the $g/\Delta$ axis, in a value relatively far from 1, is attributed to a combination of finite-size effects and the limited accuracy of the spectral entropy in higher frequencies due to the restricted sampling rate of the OTOC in time due to limited computational resources. For the case of beating signal we expect that the dominant oscillations are going to have low frequencies, since expressing the OTOC Eq.~\eqref{OTOCDef} in the eigenbasis reveals that the oscillations are of the form $e^{i(E_i-E_j)t}$ while the number operators $\langle E_i\vert \hat{n}_k \vert E_j\rangle$ connect only states nearby in energy. Therefore the limited accuracy in high frequencies is an acceptable trade-off. We also report that the spectral entropy behaves similarly for other system sizes.

Finally, the strongest indication for thermalization is the scaling of the long-time average of the OTOC
\begin{equation}\label{OTOCLongTimeAverage}
    \overline{\mathcal{C}_{ij}} = \lim_{T\rightarrow\infty}\frac{1}{T}\int_{0}^{T}dt\ \mathcal{C}_{ij}(t),
\end{equation}
with system size. For an ergodic system this late time average should scale with the system size \cite{Kukuljan2017} since the entirety of the Hilbert space is accesible with scrambling. On the contrary for a system that is MBL or scarred the system explores only some local degrees of freedom in proximity to the initial perturbation and therefore converges to some value dependent on some localization length, independent of system size \cite{RongQiang2017}. 

In Fig.~\ref{fig:OTOC_Scaling_and_Entropy}(b) we present the long-time average values for the OTOC as a function of the coupling strength $g$ for various system sizes. The system sizes we consider are all the possible sizes with $n_a \leq 3$ and Hilbert space dimension $70 \leq d_\mathcal{H} \leq 1386$. The reason to restrict $n_a\leq 3$ is because already for $n_a = 4$ only systems with very few total number of modes are accessible with exact diagonalization, which could skew the scaling analysis. Additionally, we only scaled the $\mathcal{C}_{ab}$ OTOC since it corresponds to a fixed mode distance, thus avoiding the additional scaling parameter with distance, which is not relevant for our investigation about thermalization. The way we extrapolate the scaling is with a data collapse using the scaled variable $n_a^{1/\nu_a} N_m^{1/\nu_m}(g - g_c)/\Delta$ and fit for the parameters $g_c, \nu_a$ and $\nu_m$ that yield the best collapse of all the data. The use of this dual scaling with the total number of polaritonic excitations $n_a$ and the total number of modes $N_m$ is motivated by the non-trivial scaling of the Hilbert space dimension with both of these parameters. 

The obtained fitting values are presented in the collapsed plot of Fig.~\ref{fig:OTOC_Scaling_and_Entropy}(c) and it is highly suggestive that the late-time average OTOC scales linearly with $N_m$ and with the cube of $n_a$. These sharp values suggest a precise law, possibly the linearity with the number of modes is due to the quasi-1D interactions due to the exponential amplitudes in the interacting term while the cube scaling for $n_a$ is due to the 3-body interactions. Unfortunately we are unable to provide a more precise justification for this scaling at this time. Regardless, this is strong evidence of thermalization for the system and it completes the picture for the dynamics of our system.

\section{Hilbert Space Fragmentation}

So far the dominant trend seems to be that the system undergoes a phase transition from localized to ergodic with the coupling strength playing the role of the control parameter. All the dynamical probes support this finding and the IPR of the eigenstates as well. The spectral statistics as presented through the KL divergence seem to suggest however that the system is integrable throughout this phase transition, also hinted by the slow logarithmic equilibration time of the four-point correlator of the OTOC. The question is therefore how can we reconcile these two contradicting observations and what would a potential hidden integral of motion be exactly?

\begin{figure*}
    \centering
    \includegraphics[width=0.99\textwidth]{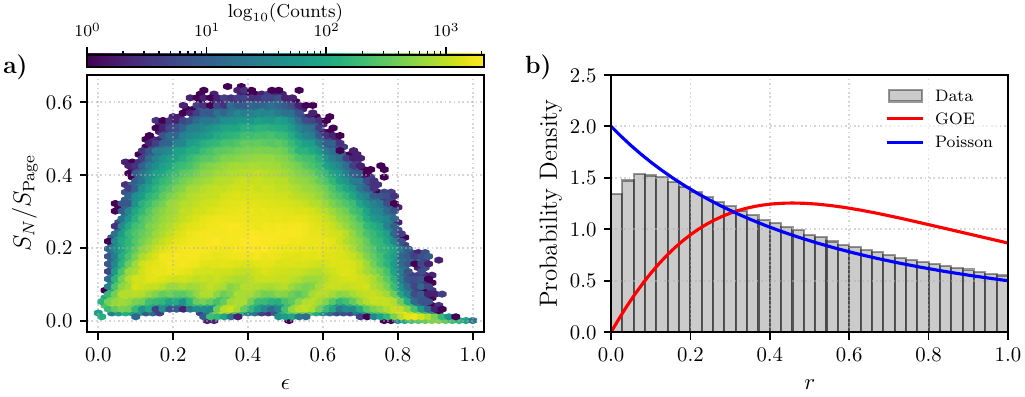}
    \caption{Energy-resolved eigenstate entanglement and spectral level spacing statistics. (a) Density distribution of von Neumann entropy $S_N$ in units of Page entropy $S_\text{Page}$ for all eigenstates across normalized energies $\epsilon = (E - E_\text{min})/(E_\text{max} - E_\text{min})$, accumulated over 200 disorder realizations. Data are obtained via exact diagonalization of Eq.~\eqref{FullHamiltonian} for a system with 8 total modes (6 polaritonic, 2 bare), $n_a = 3$ polaritonic excitations ($d_\mathcal{H} = 7168$), localization length $\xi = 1$, and coupling $g = 0.7\Delta$. Entanglement entropy is computed across a bipartition between polaritonic and bare modes. (b) Adjacent level-spacing ratio distribution (grey histogram) compared against analytical predictions for Poisson (blue) and GOE statistics (red). Data are obtained via exact diagonalization of Eq.~\eqref{FullHamiltonian} for a system with total mode count of 8 (6 polaritonic, 2 bare), total polaritonic excitations $n_a = 3$ ($d_{\mathcal{H}} = 7168$), disorder strength $w = 0.5$, interaction localization length $\xi = 1$, and coupling strength $g = 0.7\Delta$, accumulated over 200 realizations.}
    \label{fig:HSF_Evidence_Distributions}
\end{figure*}

To explore these spectral properties further, we shift our focus from global measures like the KL divergence to a direct examination of the $r$-ratio distributions [see Eqs.~\eqref{rRatiosGOEDis} and \eqref{rRatiosPoissonDis}], alongside the energy-resolved entanglement entropy between the polaritonic and bare degrees of freedom. These results are compiled in Fig.~\ref{fig:HSF_Evidence_Distributions}. The entanglement of the eigenstates in particular in Fig.~\ref{fig:HSF_Evidence_Distributions}(a) is a paradigmatic example of Hilbert space fragmentation \cite{Wang2025,Serbyn2021}, the entanglement content of eigenstates with very similar eigenenergies varies remarkably. Notice additionally that we reported the entanglement entropy in units of Page entropy $S_\text{Page} = \ln d_a - \frac{d_a}{2d_b}$ \cite{Page1993,SanchezRuiz1995}, the typical entanglement entropy that a Haar-random pure state has in a Hilbert space that has been partitioned in two subspaces of dimensions $d_a$ and $d_b$, here, referring to the polaritonic and bare modes, respectively. We deliberately prioritize this benchmark over the absolute geometric maximum of $S_\text{G} = \ln d_a$ to emphasize that the highest-entangled eigenstates fail to cross, or even reach, the typical Page curve. The entire spectrum has strictly sub-typical entanglement content, a feature fundamentally inconsistent with the standard predictions of ETH \cite{Garrison2018} and indicates that the system either has an emergent symmetry \cite{Kantaro2025} or that it is kinetically constrained \cite{Scherg2021,Pancotti2020}. In systems with an energy gradient $\Delta$ like ours, the common emergent symmetry is a dipole moment \cite{Sala2020} of the form $\sum_{n=1}^{N_b}n b_n^{\dagger}b_n + \sum_{m=1}^{N_a}(m+N_b) a_m^{\dagger}a_m$. However, this is an emergent symmetry for steep gradients (or equivalently weak coupling), while our system seems to be fragmented even for $g\approx\Delta$. This leaves the kinetic constraint as a viable candidate.

The reason the spectrum appears globally as Poisson even when the system exhibits clear signs of ergodicity can be understood in the context of general random matrix theory, recently explored in \cite{Giraud2022}. In essence, for a quantum system that is HSF, even if its independent or weakly coupled sectors have GOE spectral statistics locally, the global spectrum can appear Poisson as long as it is highly fragmented. In Fig.~\ref{fig:HSF_Evidence_Distributions}(b) we see that although the $r$-ratio distribution agrees mostly with Poissonian statistics, there is a slight dip for $r\approx 0$. This is reminiscent of the results presented in \cite{Giraud2022} for a fragmented system with local GOE spectra. Fitting our data to the analytically derived expression for the composite distribution in \cite{Giraud2022} is a computationally demanding task, however it is straightforward to check that regardless of the precise structural distribution of the fragmented sectors, the different sectors are not of equal size.

\begin{figure*}
    \centering
    \includegraphics[width=0.99\textwidth]{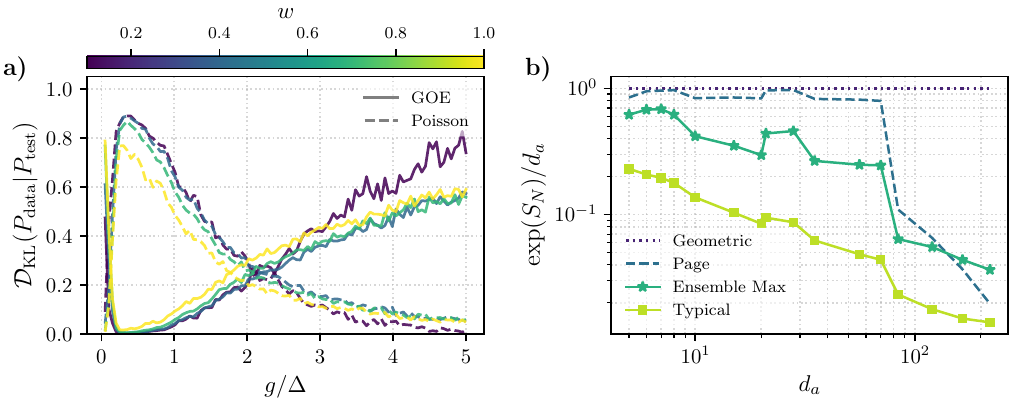}
    \caption{KL divergence for the $\xi\rightarrow\infty$ limit and Krylov scaling. (a) Kullback-Leibler divergence $\mathcal{D}_{\text{KL}}(P_{\text{data}} \vert P_{\text{test}})$ relative to GOE (solid) and Poisson (dashed) distributions as a function of normalized coupling strength $g/\Delta$. The system parameters are the exact same as in Fig.~\ref{fig:SpectralData}, but evaluated at $\xi = 10000$. (b) Effective Krylov block dimension fraction $\exp(S_N)/d_a$ as a function of polaritonic sector dimension $d_a$ (log-log scale) for $\xi = 1$, $w = 0.5$, and $g = 0.7\Delta$. Configurations span total mode numbers from 5 to 11 and excitation numbers $n_a \in [1, 9]$ ($d_\mathcal{H} \in [70, 12740]$). To balance computational cost with quantum typicality, the number of disordered realizations $N_r$ is dynamically scaled to maintain $d_\mathcal{H} \cdot N_r \approx 1.4\times 10^6$. The ensemble maximum curve tracks the single highest $S_N$ value across the ensemble, while the typical curve tracks the mean entropy of states residing in a narrow energy window ($\Delta\epsilon = 0.1$) centered at the maximum density of states in the infinite-temperature bulk.}
    \label{fig:HSF_Evidence_Scaling}
\end{figure*}

So the question arises what could this kinematic constraint be? While an exact closed-form derivation of the constraint remains analytically elusive, it has become clear to us that it is related to the exponential coefficients of the interacting term in Eq.~\eqref{IntHamiltonian}. We recall from Sec.~\ref{sec:model-characteristics} that in the limit of $\xi\rightarrow 0$ our system becomes exactly integrable, diagonalized by the displacement operator in Eq.~\eqref{DisplacementOperator}. Pushing this to the opposite extreme, $\xi\rightarrow\infty$, reveals that the effects of the HSF are vanishing and the KL divergence is finally consistent with the other figures of merit, as it is evident from Fig.~\ref{fig:HSF_Evidence_Scaling}(a). Remarkably, the system seems to be re-entrant in the Poissonian regime for very strong coupling. The nature of this re-entry is, however, beyond the scope of our original research question and will be tackled elsewhere.

Having established the presence of fragmented sectors for a fixed system size, we now address how these constraints behave as the Hilbert space expands. To quantitatively bound the dimensions of these isolated, unequal Krylov blocks across the spectrum, we track the scaling of the eigenstate entanglement metrics as a function of the polaritonic sector dimension $d_a$. To map the physical space occupied by these sectors, we plot the raw dimensional fraction $\exp(S_N)/d_a$ in Fig.~\ref{fig:HSF_Evidence_Scaling}(b). Crucially, by defining an effective subspace size via $S_N \equiv \ln d_{\text{eff}}$, effectively treating the state as if it were maximally entangled within that subspace, the quantity $d_{\text{eff}}/d_a$ serves as a strict lower bound on the relative size of the true Krylov blocks, $d_{\mathcal{K}}/d_a$. In Fig.~\ref{fig:HSF_Evidence_Scaling}(b), the ensemble maximum curve tracks the scaling of the single largest, most highly connected outlier block found across the ensemble, while the typical curve corresponds to the mean value of the entropy in a narrow energy window around the maximum of the density of states. The latter captures the typical block scaling dominating the bulk density of states. The stark, persistent separation between these curves across all dimensions visually confirms the severe size asymmetry inherent to these independent sectors. This representation reveals that the lower bound of the typical blocks scale as a vanishing fraction of the overall polaritonic sector, plummeting below $2\%$ at larger dimensions. Even the largest outlier block exhibits a declining relative envelope. The discrete, vertical drops and spikes observed at nearby values of $d_a$ directly map the structural reorganization of these blocks as the dimension of the bare modes $d_b$ expands and contracts. We also note that the ensemble maximum curve follows relatively closely the shape of the Page curve, indicating a strong dependence on the dimensions of the bare mode subspace $d_b$ while the typical curve is match smoother and appears largely unaffected by $d_b$.

Strictly speaking, the scaling analysis for the Krylov blocks in Fig.~\ref{fig:HSF_Evidence_Scaling}(b) concerns lower bounds. In isolation, this would constitute a weak statement regarding the true physical scaling of the matrix blocks $d_{\mathcal{K}}$. However, when contextualized alongside our dynamical observables, all of which demonstrate thermalization, combined with the recovery of a global GOE spectrum as $\xi \rightarrow \infty$, this provides compelling evidence that the individual Krylov blocks host independent GOE spectra. Under these ergodic sector conditions, this lower bound should track the actual scaling of the subspace dimensions closely. Ultimately, this scaling analysis demonstrates that while the absolute dimensions of the individual Krylov blocks grow with system size, they do so with a power law $d_{\mathcal{K}}\sim d_a^{\nu_a}$ with $\nu_a < 1$, establishing a strong form of Hilbert space fragmentation that is likely to survive the thermodynamic limit.

\section{Discussion}

Through the previous analysis, a clear and consistent picture has emerged about the unconventional thermalization of the three-body-mixing model under consideration. The system is HSF in kinetically constrained sectors that are ergodic. Probing the global spectrum of the system with the KL divergence of the $r$-ratio distribution yields a Poisson spectrum, because the kinetically constrained sectors do \emph{not} have level repulsions between them. This is eventually alleviated with increasing the localization length $\xi$ of the interactions by lifting the kinetic constraints and recovering global GOE spectrum. The HSF is also evidenced by the entanglement entropy of the eigenstates which varies remarkably even for eigenstates with similar energy.

Initializing the system in a far-from-equilibrium state inside one of these sectors, allows the imbalance to reach the thermal expectation value within that sector. On the other hand, the OTOC, as a basis-independent operator probe, captures both of these behaviors. In the early time regime we see an incredibly fast power-law scrambling, reflecting the fact that the initial perturbations within each sector scramble rapidly, followed by an order of magnitude slower logarithmic equilibration slowed down by the necessary communication across these kinetically constrained sectors. The late time OTOC verifies a phase transition with the coupling strength. First through the rise of spectral entropy indicating a transition of the late time oscillations from beats to thermal noise and second through the scaling of the late time average with system size, in particular a linear scaling with the total number of modes and a cubic scaling with the total number of polaritonic excitations. 

This unconventional dynamical behavior leaves several compelling open questions that warrant further theoretical exploration. First, pinning down the precise microscopic mechanism that enforces these kinetic constraints and the OTOC scaling remains an analytical priority. Furthermore, it is critical to determine how these weakly coupled sectors scale with system size and whether this fragmented structure survives the thermodynamic limit. By tracking the scaling of the eigenstate entanglement entropy, we have bounded the relative dimension of these individual Krylov blocks ($\exp(S_N)/d_a$), demonstrating that the typical sector size scales as a vanishing fraction of the total polaritonic space. To circumvent the constraints of exact diagonalization in exploring these larger system sizes and verifying the extensions of these bounds, a promising future direction lies in adapting tensor-network algorithms to track these heavily constrained dynamics. Additionally, the microscopic origin of the re-entrant Poisson transition observed at extreme coupling strengths merits closer inspection. Finally, because our current Hamiltonian is an effective model, it remains to be verified whether the full, non-truncated circuit model preserves this exact phenomenology. The latter issue places our analysis in a broader context. Effective three-body interactions have also been explored with cold atomic gases and there is convincing evidence that they appear in few nucleon systems \cite{hammer2013}. In this sense, the model (\ref{FullHamiltonian}) provides a minimal setting which carries generic features for unconventional thermalization properties.

On the experimental side, although a challenging task, the OTOC scaling is a physically accessible observable that could be measured and compared to our results. From a quantum information perspective, our findings indicate that the parameter space suitable for quantum information storage is restricted to a narrow window near zero coupling, one that systematically shrinks with increasing system size. However, the strong non-linearities of the chaotic regime present the question of whether this system can be engineered to generate highly non-linear photonic states, crucial for quantum computation with photons. Regardless, the thermalization of the three-wave mixing model exhibits novel and rich behavior and we believe it will prove an interesting testing ground for quantum thermalization. 

\section{Acknowledgments}

The authors would like to thank David DiVincenzo, Jukka Pekola, Frank Grossmann, Björn Kubala, and Ciprian Padurariu  for fruitful discussions. Financial support from the Carl-Zeiss-Stiftung (Center for Quantum Photonics), the DFG through AN336/17-1 (FOR2724), and the Ministry of Economy Baden Württemberg through KQCBW are gratefully acknowledged.

\section{Author Contributions}
E.V. and M.R. developed the code and performed the analysis and numerical calculations. J.A. coordinated the work. All authors contributed to the discussion of the results and the writing of the manuscript. Generative AI tools have been used for the proof reading of the manuscript as well as for code optimization.

\bibliography{references.bib}

\end{document}